\documentclass[twocolumn]{aa}
\usepackage{graphicx}
\usepackage{txfonts}
\begin{document}
  \title{Giant Planet Companion to 2MASSW\,J1207334$-$393254\thanks{Based on observations obtained at the Paranal Observatory, Chile, in ESO programs 73.C-0469, 274.C-5029 and 274.C-5057.}}


\author{
        G. Chauvin\inst{1}\and
        A.-M. Lagrange\inst{2}\and
        C. Dumas\inst{1}\and
        B. Zuckerman\inst{3}\and
        D. Mouillet\inst{4}\and
	I. Song\inst{3}\and
        J.-L. Beuzit\inst{2}\and
        P. Lowrance\inst{5}
}

 \offprints{Ga\"el Chauvin \email{gchauvin@eso.org}}
\institute{
$^{1}$European Southern Observatory, Casilla 19001, Santiago 19, Chile\\
$^{2}$Laboratoire d'Astrophysique, Observatoire de Grenoble, 414, Rue de la piscine, Saint-Martin d'H\`eres, France\\
$^{3}$Department of Physics \& Astronomy and Center for Astrobiology, University of California, Los Angeles, Box 951562, CA 90095-1562, USA\\             
$^{4}$Laboratoire d'Astrophysique, Observatoire Midi-Pyr\'en\'ees, Tarbes, France\\
$^{5}$Spitzer Science Center, Infrared Processing and Analysis Center, MS 220-6, Pasadena, CA 91125, USA\\
}

   \date{Received September 15, 1996; accepted March 16, 1997}

   \abstract{We report new VLT/NACO imaging observations of the young, nearby brown dwarf 2MASSW\,J1207334$-$393254 and its suggested planetary mass companion (2M1207\,b). Three epochs of VLT/NACO measurements obtained over nearly one year show that the planetary mass companion candidate shares the same proper motion and, with a high confidence level, is not a stationary background object. This result confirms the status of 2M1207\,b as of planetary mass (5 times the mass of Jupiter) and the first image of a planetary mass companion in a different system than our own. This discovery offers new perspectives for our understanding of chemical and physical properties of planetary mass objects as well as their mechanisms
of formation.}
   \maketitle
%

\section{Introduction}
During our on-going VLT/NACO survey of young, nearby
southern associations, on 27 April 2004, we observed the young brown dwarf
2MASSW\,J1207334$-$393254 (hereafter 2M1207\,A). 2M1207\,A was identified by Gizis
(2002) as a member of the 8 Myr old TW Hydrae Association (TWA), a result corroborated later by subsequent measurements at optical, infrared, and X-ray wavelengths (see references in Chauvin et al (2004) and proper motion data presented below). 

In the close circumstellar environment of 2M1207\,A, we discovered a faint planetary mass companion candidate (hereafter 2M1207\,b; see Chauvin et al. 2004). Based on HK and L$\,\!'$ photometry and evolutionary models, we found a predicted mass of 5 Jupiter masses. On 28 August 2004, Schneider et al. (2004) obtained complementary
photometric observations at 0.9, 1.1 and 1.6$\,\mu$m, using the Hubble
Space Telescope (HST). Due to the small time span between the
ground-based and space observation of 2M1207\,A, the HST results could not
unambiguously confirm the companionship of the candidate planet, but corrobate the mid-L dwarf spectral type estimated with the H-band VLT/NACO spectrum. 

In this letter, we report new VLT/NACO observations of 2M1207\,A and b, obtained on 5 February 2005
and on 31 March 2005. They clearly demonstrate that the two are
comoving and enable us to reasonably confirm the status of the giant planet\footnote{We have adopted here the \textit{International Astronomical Union} definition to differentiate planetary and brown dwarf companions.} companion to the brown dwarf 2M1207\,A.

%

\section{Observations}

2M1207\,A and b were imaged on 5 February and 31 March 2005, using the 
infrared wavefront sensor of VLT/NACO. For both epochs, a set of 12 jittered images was taken with the K$_{s}$ filter and the S27 camera of CONICA, leading to a total exposure time of $\sim12$~min on the source. The observing log is reported in Table~1. 

To calibrate the platescale and the detector orientation, we observed at each epoch the astrometric field of $\theta$ Ori 1 C. The orientations of true north of the S27 camera were found on 27 April 2004, 5 February 2005 and 31 March 2005 respectively at $+0.08^o$, $+0.07^o$ and $+0.14^o$ east of the vertical with an uncertainty of $0.10^o$. The pixel scale was found to be relatively stable in time with values of $27.01\pm0.05$\,mas. 

After cosmetic reductions using \textit{eclipse} (Devillard 1997), we used
the deconvolution algorithm of V\'eran \& Rigaut (1998) to obtain the
 position of 2M1207\,b relative to 2M1207\,A at each epoch. Image
selection of the AO observations was carefully performed and in Table 2 we slightly
revise the published separation and position angle of the 
first epoch detection (Chauvin et al 2004). 
\begin{table}[t]
\centering          
\caption{VLT/NACO observing Log of 2M1207\,A and b}
\label{tab:setup}      
\begin{tabular}{llllllll}     
\hline\hline       
UT Date       & Seeing   &  $\tau_{0}^*$   &  Airmass &  Strehl &   \footnotesize{FWHM} \\
             &(arcsec)  & (ms)          &          & (\%)    & (mas)                   \\
\noalign{\smallskip}\hline\noalign{\smallskip}
27/04/2004  & 0.52     & 14.1          & 1.08     & 23   & 81       \\
05/02/2005  & 0.69     & 7.3           & 1.18     & 18   & 92       \\
31/03/2005  & 0.86     & 6.4           & 1.10     & 15   & 95      \\
\hline                  
\end{tabular}
\begin{list}{}{}
\item[$^*$] $\tau_{0}$ corresponds to the atmospheric correlation time at 0.5~$\mu$m
\end{list}
\end{table}
\vfill
%
\section{Companionship confirmation}

\subsection{Proper motion and parallactic motion of 2M1207\,A }

Using the USNO-B1.0 Catalogues (Monet et al. 2003), Gizis (2002) gave a first estimation of the 2M1207\,A proper motion: $(\mu_{\alpha},\mu_{\delta})\sim(-100, -30)$~mas/yr. Using all available positions from the 2MASS (Cutri et al. 2003), DENIS (Epchtein et al. 1997) and SSS (Hambly et al. 2001) catalogues, Scholz et al. (2005) have recently determined a more accurate proper motion of 2M1207\,A: $(\mu_{\alpha},\mu_{\delta})\sim(-78\pm11, -24\pm9)$~mas/yr.

Parallactic motion can be determined based on the distance and the first observing epoch of 2M1207\,A. Presently, no astrometric parallax has been measured for 2M1207\,A. Using various evolutionary models (Burrows et al. 1997, D'Antona \& Mazzitelli 1997, Baraffe et al. 1998), we conservatively infer absolute magnitudes M$_{\rm{J}}=8.8\pm0.5$ and M$_{\rm{K}}=7.8\pm0.5$ for a young M8 dwarf of $\sim10$~Myr, like 2M1207\,A. Then, with apparent magnitudes $J = 13$ and $K = 11.95$ (Cutri et al. 2003), a J-band and K-band distance modulus of $4.2\pm0.5$ mag implies a distance $\sim70\pm20$~pc to 2M1207\,A. We will consider hereafter this $1~\sigma$ error bar on this distance estimate to be conservative. 

\begin{table*}[t]
\caption{Offset positions of 2M1207\,b from A on 27 April 2004, 5 February 2005 and 31 March 2005}             
\label{table:1}      
\centering          
\begin{tabular}{lllllllllll}     
\hline\hline                      
             &  \multicolumn{4}{c}{\underline{\hspace{2.4cm}OBSERVED\hspace{2.4cm}}}  &   \multicolumn{2}{c}{\underline{\hspace{0.9cm}IF BACKGROUND\hspace{0.9cm}}}    &      &           \\
UT Date      &    $\Delta\alpha$             & $\Delta\delta$              & Separation     &    PA           &    $\Delta\alpha$ (IF BKG)             & $\Delta\delta$ (IF BKG) & $\chi^2$ &   Probability\\ 
             &    (mas)                  &    (mas)                 &    (mas)       &  ($^o$)          &    (mas)                 &    (mas)   &   &         ``B is BKG''     \\
 \noalign{\smallskip}\hline\noalign{\smallskip} 
27/04/2004  &$629\pm4$         & $-447\pm5$       & $772\pm4$      &$125.4\pm0.3$    & -  &- & -&-     \\
05/02/2005  &$626\pm5$         & $-446\pm4$       & $768\pm5$      &$125.4\pm0.3$    & $673\pm12$   & $-624\pm10$    &  20&$4$\,e-5\\
31/03/2005  &$631\pm7$         & $-451\pm10$      & $776\pm8$      &$125.5\pm0.3$    & $695\pm13$   & $-421\pm14$    &  49&$<1\,$e-9\\
\hline                  
\end{tabular}
\end{table*}

\begin{figure}[t]
\centering
\includegraphics[width=8.5cm]{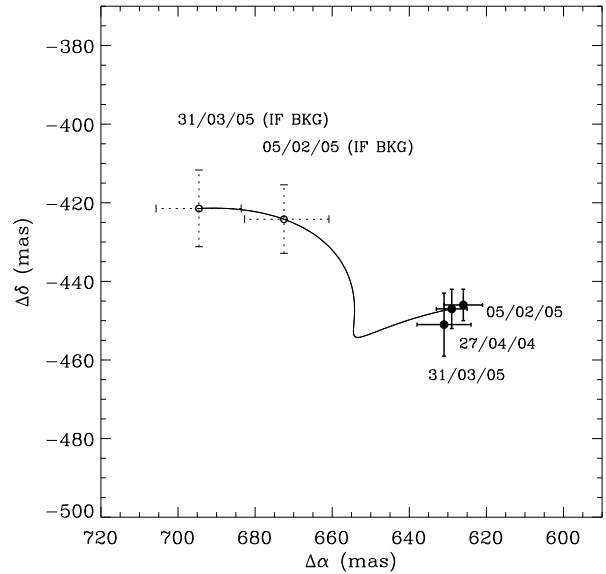}
\caption{VLT/NACO Measurements (3 \textit{full circles} with uncertainties) of the offset positions of 2M1207\,b relative to A, obtained on 27 April 2004, 5 February 2005 and 31 March 2005. The expected variation of offset positions, if b is a background object, is shown (\textit{solid line}), based on a distance of $70$~pc, a proper motion of $(\mu_{\alpha},\mu_{\delta})=(-78, -24)$~mas/yr for A and the initial offset position of b from A. The 2 \textit{empty circles} give the corresponding expected offset positions of a background object on 5 February 2005 and 31 March 2005, with total uncertainties, which are the quadratic sum of individual uncertainties (distance, proper motion of A and initial position of b from A).}
\label{Fig:image}
\end{figure}

\subsection{2M1207\,b is comoving}

To verify that 2M1207\,A and b were comoving together in the sky, we measured their relative position on 27 April 2004, on 5 February and on 31 March 2005. The separations and position angles are reported in Table~2 and the corresponding uncertainties, at each epoch, include errors from the deconvolution process and from the detector calibrations (platescale and detector orientation, see section~2). 

Fig.~1 displays, in a ($\Delta\alpha$, $\Delta\delta$) diagram, the offset positions of 2M1207\,b from A, observed with NACO on 27 April 2004, 5 February 2005 and 31 March 2005. The expected evolution of the relative A-b positions, under the assumption that b is a stationnary background object, is indicated for 5 February and 31 March 2005 in Fig.~1 (2 \textit{empty circles}) and in Table 2.  The corresponding estimations are based on the first measurement on 27 April 2004, and on the expected proper motion and parallactic motion of A (see section 3.1). As each of these three parameters are independent, the total uncertainty is the quadratic sum of individual uncetainties. Note that, if bound, the orbital motion of b around A would be lower than 2 mas/yr (negligible). 

\begin{figure}[t]
\centering
\includegraphics[width=8.5cm]{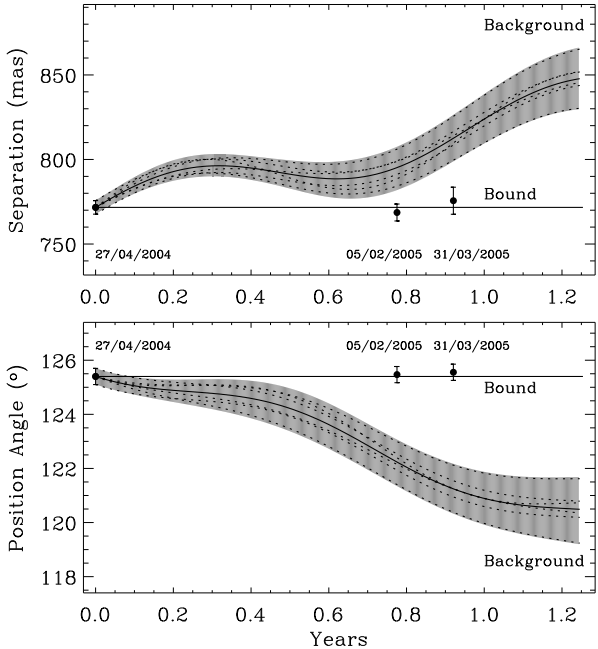}
\caption{Offset positions in terms of separation (\textit{Top}) and position angle (\textit{Bottom}) of 2M1207\,b from  A, in 27 April 2004, 5 February 2005 and 31 March 2005 (\textit{full circles} with uncertainties). The expected variation of offset positions, if b is a background object, is shown (\textit{solid line}), based on a distance of $70$~pc, a proper motion of $(\mu_{\alpha},\mu_{\delta})=(-78, -24)$~mas/yr for A and the initial offset position of b from A. We have also reported the associated uncertainties (\textit{shaded region}) and the different contributions (distance, proper motion, initial offset position) in \textit{dotted lines}.}
\label{Fig:image}
\end{figure} 

Fig.~2 shows the observed offset position of 2M1207\,b from A at each epoch, decomposed in terms of separation (\textit{Top}) and position angle (\textit{Bottom}). The given uncertainties are taken from Table~2. As for Fig.~1, we show the expected variation of offset positions, if 2M1207\,b is a background object. 

Using Fig.~1, we can statistically test the possibility that 2M1207\,b is a background object. Under that assumption, the expected offset positions (2 \textit{open circles} on Fig.~1) and measured offset positions (3 \textit{full circles}) would coincide within uncertainties (including for each epoch, the uncertainties for the measurement at that epoch, and the total uncertainty on the expected offset position if background, see Table~2). The normalized $\chi^2$ of 4 degrees of freedom (corresponding to the 4 measurements: separations in the $\Delta\alpha$ and $\Delta\delta$ directions for 2$^{\rm{nd}}$ and 3$^{\rm{rd}}$ epochs) is 49. Consequently, the corresponding probability that 2M1207\,b is a background star is less than 1e-9.

In Chauvin et al (2004), we estimated the probability of finding a
field L-dwarf within 780~mas of 2M1207\,A to be less than one in a million.
From data, given in Vrba et al (2004), we deduce that a fraction of old
late L-type field brown dwarfs of the same apparent magnitude as 2M1207\,b
would have the same proper motion as 2M1207\,b.  However the
probability that the proper motion direction of a field dwarf would so
nearly mimic that of 2M1207\,A is small.  The net product of all these
probabilities is impressively tiny and, therefore, thanks to NACO at the
VLT, we can reasonably claim that 2M1207\,A and b are physical companions.
Continued monitoring of A and b will further solidify their relationship.

%
\section{Discussion}

\subsection{Mass of 2M1207\,b}
 
In the 2M1207\,b discovery paper (Chauvin et al 2004), we relied on
evolutionary models of Burrows et al. (1997), Chabrier et al. (2000) and Baraffe et al. (2002) to derive a mass of 5~M$_{\rm{Jup}}$ for
2M1207\,b (see Table~3).  However, such models, when applied to certify the mass of young
substellar objects, remain essentially untested in an absolute sense.
Masses of substellar objects can also be derived through model independent
techniques.  In two thoughtful papers, Mohanty et al. (2004a \& b) use
synthetic spectra to derive the gravity and then the masses of substellar
objects in the $\sim5$~Myr old Upper Scorpius association.  Their analysis
suggests, but does not prove, that for masses $\le30$\,M$_{\rm{Jup}}$ the
evolutionary models overestimate the masses of the Upper Sco members,
while for high mass brown dwarfs just the opposite is true.  (In the high
mass brown dwarf regime, the Mohanty et al conclusion agrees qualitatively
with that of Close et al (2005) who analyzed the $\sim50$~Myr old close binary
AB Dor).

Because the Upper Sco and TW Hya associations are of similar age, if the
Mohanty et al (2004a \& b) analysis is correct, then the mass of 2M1207\,b could be actually
smaller than the $\sim5$\,M$_{\rm{Jup}}$ we derived using evolutionary models.

\subsection{Origin of 2M1207\,b}

The system 2M1207\,A and b is roughly characterized by a mass ratio $q\sim0.2$ and semi-major axis $\geq55$~AU (at
70~pc). 2M1207\,A and b is very different from known brown dwarf binaries
discovered in the field (Close et al. 2003, Gizis et al. 2003, Burgasser
et al. 2003, Bouy et al. 2003) and within star forming regions
(Neuh\"auser et al. 2002), which are characterized by semi-major axes
$<15$~AU and a trend toward equal-mass components. The primary 2M1207\,A harbours an accreting disk, revealed by
the strong H$\alpha$ emission line (Gizis 2002, Mohanty et al.
2003), the non-chromospheric X-ray activity (Gizis \& Bharat 2004) and
mid-IR excess emission at 8.7~$\mu$m and 10.4~$\mu$m (Sterzik et al.
2004). 

We can then wonder if 2M1207\,b has formed within the circumstellar disk of 2M1207\,A via a two-step process that begins with core accretion (Pollack et al. 1996) or via gravitational collapse induced by disk instability. This origin would imply the existence, at an earlier phase, of a disk around 2M1207\,A with a mass of at least 20\% of the primary mass in order to form the planetary mass companion. This is very unlikely in view of our current knowledge of circumstellar disk masses around young stars, which hardly exceed 1-10\% of the primary mass. 

This binary has then probably formed via gravitational collapse processes. We can inspect the four main mechanisms, summarized by Kroupa \& Bouvier (2003) and presented below, to see if they might account for the origin of 2M1207\,b (the first three are actually a gravitational collapse terminated by a loss of the accretion envelop):
\begin{itemize}
\item the ``embryo-ejection'' model, which can be excluded, as the maximal separation predicted for the widest brown dwarf binaries is about a$_{\rm{max}}\sim10-20$\,AU (Reipurth \& Clarke 2001, Bates et al. 2002),
\item the ``collision'' model, excluded too, as it would have led to a disruption of the young binary system,
\item the ``photo-evaporation'' mechanism within a massive star formation association. This might be possible as TWA is possibly related to the sub-region Lower Centaurus-Crux of the recent massive star forming region Sco-Cen (Mamajek \& Fiegelson 2001; Mamajek et al. 2002; Song et al. 2003). About 10\,Myr ago, a few O-type stars could have blown up as supernovae in this sub-region and played a role in both the formation and the dispersal of young, nearby stars, including 2M1207\,A and b.

\item Finally, the ``star-like'' mechanism, i.e a gravitational collapse without loss of accretion envelop. The mass of 2M1207\,b falls in the vicinity of the minimum mass fragments proposed by Low \& Lynden-Bell (1976), who paint a picture of Jeans mass fragmentation in a dark molecular cloud. They derive a minimum mass of $\sim7$\,M$_{\rm{Jup}}$, a robust result which is insensitive to changes in gas opacity and to dust and cosmic ray characteristics.
\end{itemize}
\begin{table}[t]
\label{table:1}      
\centering          
\caption{Physical characteristics of 2M1207 A and b according to DUSTY models of Chabrier et al. (2000) and Baraffe et al. (2002)}
\begin{tabular}{llllllll}     
\hline\hline                      
Name		        & Age    &Mass	                &T$_{\rm{eff}}$	&Luminosity    & Radius\\
	     	        &(Myr)   &(M$_{\rm{Jup}}$)	&(K)            &log(L/L$_{\odot}$)   & (R/R$_{\odot})$\\
\noalign{\smallskip}\hline\noalign{\smallskip} 
2M1207\,A	        & 5       &$25$	                &$2600$	&$-2.43$               &$0.31$            \\
	                & 10       &$45$	                &$2800$	&$-2.36$               &$0.28$            \\
2M1207\,b	        & 5       &$4$                    &$1163$      &$-4.28$     &$0.15$    \\ 
	                & 10       &$6$                    &$1320$      &$-4.24$     &$0.15$    \\ 
\hline                  
\end{tabular}
\end{table}

\subsection{Perspectives of direct detections}

An ultimate breakthrough in modern astronomy will be to image and study extrasolar planets down to the mass of the Earth and determine if they contain signs of life. Our imaging of a 5 Jupiter mass planet is a first step in this direction. Modern instrumentation will soon permit astronomers to carry out a detailed study of the physical and chemical characteristics of extra-solar planets.

In addition, TWA is at an age ($\sim$10\, Myr) that is believed to have been a critical time during the formation of our solar system. For example, according to measurements of isotopes of hafnium-tungsten in meteorites and in the mantle, it is deduced that Earth's core likely formed in less than 30 million years (see Zuckerman \& Song 2004). Thus, by studying various properties of the circumstellar vicinity of stars and brown dwarfs of order tens of millions of years of age, we can learn things about our own origins that are difficult or impossible to decipher from investigation of old, mature planetary systems.

%

\section{Conclusion}

We have presented new high contrast images and proper motion measurements of the young brown dwarf 2MASSW\,J1207334$-$393254, a member of the TW Hydrae association. The images confirm, at a high confidence level, that the brown dwarf has a companion whose mass is about 5 times the mass of Jupiter and perhaps even less (see Section 4.1) and is, therefore, the first image of a planetary mass companion in a system other than our own. It is very unlikely that this giant planet formed within a circumstellar disk, but more probably via one of two formation mechanisms proposed for brown dwarfs. This discovery offers proof that such mechanisms can form bodies down to the planetary mass regime.

\begin{acknowledgements}
We thank N. Ageorges, C. Lidman, D. Nuernberger, S. Mengel and L. Tacconi-German from ESO for carrying out our observations in service mode, as well as G. Chabrier, F. Crifo, S. Metchev, M. Sterzik. for helpful discussions. Finally, we thank T. Fusco for important inputs provided for the deconvolution of 2M1207\,A and b and referee John Gizis for helpful suggestions.
\end{acknowledgements}

\end{document}